\documentclass[aps,prappl,reprint,longbibliography,superscriptaddress]{revtex4-1}  
\pdfoutput=1 
\usepackage{graphicx}  
\usepackage{dcolumn}   
\usepackage{multirow}
\usepackage{bm}        
\usepackage{amssymb}   
\usepackage{amsmath}   
\usepackage{siunitx}   
\usepackage{color}     
\usepackage{fdsymbol}
\usepackage{dcolumn,booktabs}

\usepackage{blindtext}

\usepackage[version=4]{mhchem}
\usepackage{float}
\usepackage{verbatim}
\usepackage[normalem]{ulem} 
\usepackage{subcaption}
\usepackage{ragged2e}
\DeclareCaptionJustification{justified}{\justifying}
\usepackage{soul}
\usepackage{xcolor}

\DeclareSIUnit\bar{bar}

\begin{document}


\title{Mitigation of interfacial dielectric loss in aluminum-on-silicon superconducting qubits}

\author{Janka~Biznárová}

\author{Amr~Osman}
\author{Emil~Rehnman}
\author{Lert~Chayanun}
\author{Christian~Križan}

\affiliation{Chalmers University of Technology, Microtechnology and Nanoscience, SE-41296, Gothenburg, Sweden}

\author{Per~Malmberg}
\affiliation{Chalmers University of Technology, Chemistry and Chemical Engineering, SE-41296, Gothenburg, Sweden}

\author{Marcus~Rommel}
\author{Christopher~Warren}
\author{Per~Delsing}
\author{August~Yurgens}
\author{Jonas~Bylander}
\author{Anita~Fadavi~Roudsari}

\affiliation{Chalmers University of Technology, Microtechnology and Nanoscience, SE-41296, Gothenburg, Sweden}
\vskip 0.25cm

\date{\today}

\sisetup{range-phrase=--}

\begin{abstract}
We demonstrate aluminum-on-silicon planar transmon qubits with time-averaged $T_1$ energy relaxation times of up to 270~$\mu$s, corresponding to $Q=5$~million, and a highest observed value of 501~$\mu$s. We use materials analysis techniques and numerical simulations to investigate the dominant sources of energy loss, and devise and demonstrate a strategy towards mitigating them.
The mitigation of loss is achieved by reducing the presence of oxide, a known host of defects, near the substrate-metal interface, by growing aluminum films thicker than 300~nm. 
A loss analysis of coplanar-waveguide resonators shows that the improvement is owing to a reduction of dielectric loss due to two-level system defects. 
We perform time-of-flight secondary ion mass spectrometry and observe a reduced presence of oxygen at the substrate-metal interface for the thicker films. 
The correlation between the enhanced performance and the film thickness is due to the tendency of aluminum to grow in columnar structures of parallel grain boundaries, where the size of the grain depends on the film thickness: transmission electron microscopy imaging shows that the thicker film has larger grains and consequently fewer grain boundaries containing oxide near this interface.
These conclusions are supported by numerical simulations of the different loss contributions in the device.
\end{abstract}


\maketitle


\section{Introduction}

Limited qubit coherence is still one of the main challenges for developers of solid-state quantum computing hardware. In the gate model of quantum computation, errors per quantum gate well below a tenth of a percent are required to execute meaningful quantum algorithms---and the lower the errors, the less overhead is needed for error correction and error mitigation.

Superconducting qubits represent a leading platform for quantum computation. Their coherence time improvement, from nanoseconds at their conception to milliseconds today, is nothing less than remarkable \cite{Nakamura1999, Houck2008, Kjaergaard2020, Somoroff2023}. Progress has been driven by discoveries and systematic engineering to identify and mitigate the sources of decoherence. This effort has two components: a reduction of the qubits’ sensitivity to noise by development of innovative device design concepts  \cite{Koch2007, Sage2011,dunsworth2017,Gordon2022}, and a reduction of the noise itself by engineering of the qubits’ electromagnetic environment \cite{Burnett2019,Gordon2022}, improved fabrication methods \cite{Megrant2012,Nersisyan2019}, and materials science {\cite{Chang2013,dunsworth2017,Murray2021}. 

Coplanar-waveguide (CPW) resonators and transmon qubits are commonly used for studying coherence and identifying loss mechanisms  in superconducting devices. The dominating decoherence source  within these devices is dielectric loss and noise, attributed to charged two-level-system (TLS) defects  at surfaces and interfaces, i.e., in thin sheets at the substrate-metal (SM), metal-air (MA), and substrate-air (SA) interfaces {\cite{Martinis2005,macha2010,Lisenfeld2019,molina,niepce2021, Osman2023}. Oxide-based defects are known contributors to TLS loss {\cite{lisenfeld2015,Muellertls}. 

Among superconducting materials for qubits, aluminum (Al) has been  a dominant material within the community: the ease and reproducibility of thin-film deposition, low cost, well-described chemistry, and the robust Josephson effect in small tunnel junctions comprised of aluminum oxide sandwiched between aluminum electrodes have made this material an attractive choice. 

The achieved energy relaxation time ($T_1$) of transmon qubits in a planar geometry whose wiring layer is made of aluminum is currently about \SI{100}{\micro\second} \cite{dunsworth2017}. However, recent years have seen an impressive improvement for transmon qubits with wiring layers made of tantalum, a less explored material for quantum devices, showing average $T_1$ in the range of 300~–~\SI{480}{\micro\second} \cite{Place2021,Wang2022}. This progress has subsequently spread to the more traditionally used materials titanium nitride and niobium, with average $T_1$ up to \SI{291}{\micro\second} and $210~\mu$s}, respectively \cite{Deng2023,Kono2023}. An apparent advantage of these materials over aluminum is their ability to withstand harsh chemical surface cleaning.

Here we demonstrate planar transmon qubits made of aluminum on silicon substrates, with time-averaged $T_1$ of up to $270~\mu$s ($Q = \omega_q T_1 = 5$~million, $\omega_q$: qubit frequency). The improvement is mirrored in CPW resonators' quality factor measurements: the Q factor’s dependence on circulating power indicates that the increased relaxation time is due to a reduction of TLS losses. 
We achieve this improvement by depositing a thicker layer of aluminum---300~nm or more---compared to our previous standard of 150~nm. We conduct material depth profiling by time-of-flight secondary ion mass spectrometry (ToF-SIMS) and identify an aluminum-film-thickness dependence of the oxygen concentration at the SM interface, which decreases  for thicker films.
Further, we show by transmission-electron microscopy (TEM) that the thicker film has larger grains; we therefore interpret the reduced oxygen presence as being due to a smaller prevalence of grain boundaries in the thicker film and hence fewer sources of interfacial dielectric loss. 
Our numerical simulations of the loss participation due to the different dielectrics present in the device support this interpretation.

\begin{figure*}
    \centering 
    \includegraphics[width=17.5cm]{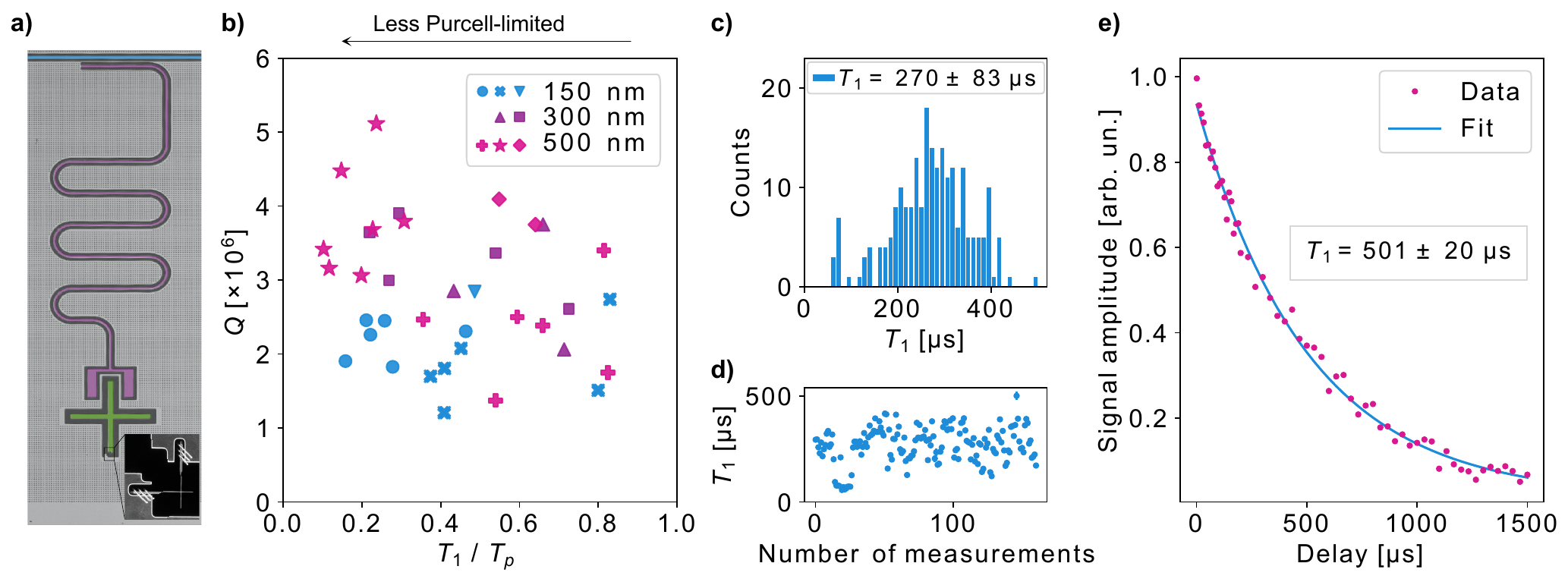}
    \caption{\label{fig:qubits}
    Qubit design and coherence data.
    \textbf{a)} False-colored micrograph of a qubit device. A portion of the input/output transmission line is shown in blue, coupled to a  readout resonator, shown in purple. The qubit capacitor is shown in green. The inset shows a scanning electron micrograph (SEM) of the Josephson junction. \textbf{b)} Time-averaged qubit quality factor $Q$ as a function of the measured qubit relaxation time relative to the calculated Purcell decay time, $T_1/T_p$, for qubits fabricated on wiring layers of 150, 300, and \SI{500}{\nano\meter}. The quality factor of those qubits with $T_1 \ll T_p$ is closer to the limit set by the TLS loss. Qubits displayed by the same marker are made on one wafer. 
    \textbf{c)} A histogram showing the relaxation time of the best qubit with average $T_1 =\,$\SI{270}{\micro\second} plus/minus one standard deviation of \SI{83}{\micro\second}. \textbf{d)} 160 $T_1$ measurements of the qubit in \textbf{(c)} over a span of \SI{48}{\hour}. Fit error bars are smaller than markers where not visible. \textbf{e)} Demonstration of the exponential fit to the longest $T_1$ measured. } 
\end{figure*}

\section{\label{sec:qubits} Qubit coherence}

We show a representative schematic of the transmon (Xmon) qubit circuit in figure~\ref{fig:qubits}(a).
The device has a 2D architecture}  and consists of a Josephson junction shunted by a large  capacitor, which is coupled to a quarter-wavelength ($\lambda$/4) CPW resonator. The CPW resonator is used for readout as well as qubit control. The devices are fabricated on silicon (Si) substrates and stripped of surface oxides prior to Al deposition. We present more details on device fabrication and measurement in sections~\ref{sup:fab} and \ref{sup:setup}, respectively.

In figure~\ref{fig:qubits}(b), we present coherence data of 36 out of the 38 measured qubits made with Al films of various thicknesses of \SI{150}{\nano\meter} (previously our standard thickness), \SI{300}{\nano\meter}, and \SI{500}{\nano\meter}. Two qubits were disregarded from the analysis due to inconsistent measurement results, see section \ref{sup:QubitsTable}. The figure shows a trend towards higher coherence times with thicker films. The average quality factor $Q$ is $\text{2.1}\times \text{10}^\text{6}$ for the qubits on the \SI{150}{\nano\meter} thick film, and $\text{3.2}\times \text{10}^\text{6}$ for those on the thicker films.

The qubit frequencies range from \SI{2.8}{\giga\hertz} to \SI{5.0}{\giga\hertz}, while the resonator frequencies are between \SI{6.0}{\giga\hertz} and \SI{6.8}{\giga\hertz}, see table~\ref{tab:qubits}. 
For simplicity of device design and characterization, there is no Purcell filtering present in the device to mitigate spontaneous emission from the qubit into its resonator and the transmission line. Therefore, for these long-$T_1$ qubits, Purcell decay is non-negligible and needs to be included among the limiting factors of the qubits' lifetime

\begin{equation}\label{eq:QubitsQ}
    \frac{1}{Q} = \frac{1}{Q_{\mathit{TLS}}}  + \frac{1}{Q_{p}} + \frac{1}{Q_{\mathit{qp}}}+ \frac{1}{Q_{\mathit{rad}}} + \ldots\,
\end{equation}
Here $Q_{\mathit{TLS}}$, $Q_{p}$, $Q_{\mathit{qp}}$, and $Q_{\mathit{rad}}$ are the theoretical quality factors attributed to TLS, Purcell decay, quasiparticles, and other radiation losses not related to Purcell decay, respectively.

The time constant  of the Purcell decay for each qubit is calculated as $T_p = \gamma^{-1}$, with the decay rate $\gamma$  \cite{Koch2007},
\begin{equation}\label{eq:PurcellDecay}
    \gamma = \kappa \frac{g^2}{\Delta^2}\,,
\end{equation}
where the cavity decay rate is $\kappa = \omega_r/Q_l$, the qubit-resonator coupling is $g = \sqrt{\chi \Delta} $, and the qubit-resonator detuning is $\Delta = \omega_q - \omega_r $. Here $\omega_r$ denotes the resonator frequency, $Q_l$ represents the resonator's loaded (total) quality factor and $\chi$ is the dispersive frequency shift. 

The qubit-resonator detunings vary between the devices due to variation in the Josephson junctions. Qubits with a smaller qubit-resonator detuning are more strongly Purcell-limited than others. As a result, a direct comparison between the qubits based solely on their $T_1$ is not the correct way of evaluating the qubits' performance.

To compare qubits of different frequencies while simultaneously taking the effect of the Purcell decay into account, we plot the qubits' quality factor $Q$ as a function of $T_1/T_p$ in figure~\ref{fig:qubits}(b). We choose $Q$ instead of $T_1$ since the latter depends directly on frequency, which also varies across devices. Now, rewriting Equation~(\ref{eq:QubitsQ}) while ignoring the loss due to quasiparticles and radiation, we are left with $(1/Q)(\text{1}-T_1/T_p) =1/Q_{\mathit{TLS}}$. The quality factor will therefore approach its limit set by the TLS loss when $T_1/T_p \ll \text{1}$.

In figure~\ref{fig:qubits}(b), the data shown in blue belong to qubits fabricated on a \SI{150}{\nano\meter} thick wiring layer, while the purple and pink markers show qubits fabricated on 300 and \SI{500}{\nano\meter} thick layers, respectively. 
While the average $Q$ for the thicker films is higher, the separation between the datasets scales with the Purcell effect. When $T_1 \leq 0.5\,T_p$, a range at which the qubits are not significantly limited by Purcell decay, the data for the thicker and thinner films are clearly separated, with the qubits fabricated with the thinner film (\SI{150}{\nano\meter})  showing a lower quality factor compared with those on the thicker films ($\geq$~\SI{300}{\nano\meter}). There is a subtle difference between the quality factors of the qubits on 300 and \SI{500}{\nano\meter} films; however, the difference is not so pronounced as to clearly differentiate the qubits with these two film thicknesses. The average quality factor in this range is $\text{2.1}\times \text{10}^\text{6}$ for the qubits on the thin film, while it increases by 66\,\% to $\text{3.5}\times \text{10}^\text{6}$ for those on the thicker films. 
For a more stringent case of $T_1 \leq 0.25\,T_p$, the average quality factor becomes $\text{3.8}\times \text{10}^\text{6}$ for the thicker films. Assuming a typical qubit frequency of $\omega_q/\text{2}\pi$ = \SI{3}{\giga\hertz}, the corresponding relaxation time becomes $\sim$\SI{200}{\micro\second} for a qubit made from the thicker 300 - \SI{500}{\nano\meter} films, in the weakly Purcell-limited regime.  

We remark that the coherence times quoted in figure~\ref{fig:qubits}(b) are mean values gathered by measuring each qubit for at least \SI{24}{\hour}. This is necessary to accurately assess a qubit's quality, since the coherence time of superconducting qubits is known to fluctuate over time \cite{Chang2013,Muellertls,Burnett2019,Schlor2019}. In figure~\ref{fig:qubits}(c) we show the histogram of the relaxation time of the best qubit, labeled \textit{Q27} in table~\ref{tab:qubits}. The histogram contains 160 averaged values (see figure~\ref{fig:qubits}(d)) obtained over a span of \SI{48}{\hour}. We find an average relaxation time of \SI{270}{\micro\second}, on par with the numbers reported for the best qubits on titanium nitride and niobium films \cite{ Kono2023, Deng2023}. In figure~\ref{fig:qubits}(e) we present the longest measured relaxation time for the aforementioned qubit, together with its corresponding fit, showing $T_1 = $ \SI{501}{\micro\second}.

The improvement of the relaxation time in the thicker films is also reflected in the values obtained for the spin echo decoherence time $T_2\,^\text{echo}$, as $T_2\,^\text{echo} > T_1$ for most of our qubits, see table~\ref{tab:qubits}. Although we find mean values as long as \SI{307}{\micro\second}, the spin echo decoherence time does not quite reach the theoretical maximum of $2\,T_1$, indicating that the coherence of our qubits is not solely limited by $T_1$, or by the type of dephasing noise typically cancelled by the spin echo sequence.

\section{\label{sec:resonators} Resonator quality}
In order to distinguish and quantify the particular energy loss mechanisms at play, we also fabricate and measure 50 bare $\lambda$/4 CPW resonators with resonance frequencies in the range of 4 -- \SI{8}{\giga\hertz}. The internal quality factors $Q_i$ of the resonators are extracted by fitting the transmission scattering parameter $S_{21}$ vs.\@ frequency using a resonance circle fitting method with diameter correction adapted directly from \cite{probst2015}.

For TLS-loss-limited resonators, $Q_i$ at low drive powers will be diminished according to the model for interacting TLSs \cite{gao_phd,Pappas2011,goetz2016},
\begin{equation}\label{eq:QiFit}
    \frac{1}{Q_i } = F \delta^0_{\mathit{TLS}} \frac{\tanh{(\hbar \omega_r / 2 k_B T)}}{\left( 1 + \langle n \rangle / n_c\right)^\beta} + \delta_0\,,
\end{equation}

\noindent where $F \delta^0_{\mathit{TLS}}$ describes the loss due to parasitic TLSs, with $F$ quantifying the TLS filling (also known as the total TLS participation ratio), and $\delta^0_{\mathit{TLS}}$ the TLS density. $n_c$ is the critical photon number required to saturate a single TLS on average, and $\beta$ describes how quickly the TLSs
saturate with power. The temperature-dependent factor $\tanh{(\hbar \omega_r / 2 k_B T)}$ is $\sim$1 in our temperature regime ($T$ denotes the temperature, and $\hbar$ and $k_B$ are the Planck and Boltzmann constants, respectively).
The power-independent term $\delta_0$ quantifies other sources of loss, such as resistive or radiative losses. The average number of photons circulating in the resonator $\langle n \rangle$ is estimated using
\begin{equation}\label{eq:nPhoton}
    \langle n \rangle = 2\frac{Z_0}{Z_r}\frac{Q_l^2}{Q_c}\frac{P_{\mathit{in}}}{\hbar\omega_r^2}\,,
\end{equation}
\noindent where $Z_0$ and $Z_r$ are the characteristic impedances of the transmission line and the resonator, respectively, and  $P_{\mathit{in}}$ is the microwave power delivered to the input port of the device. $Q_l$ and $Q_c$ are the loaded and coupling quality factors, respectively.

In figure~\ref{fig:resonators}(a), we compare the extracted $Q_i$ of two resonators with the same resonance frequency ($\omega_r/2\pi =\,$ \SI{4.45}{\giga\hertz}) on 150 and \SI{500}{\nano\meter} films. Both resonators show a similar $Q_i$ when $\langle n \rangle\gtrsim$ \text{$10^6$}. However, the resonator fabricated on the \SI{500}{\nano\meter} film has a higher internal quality factor at lower photon levels. We fit the quality factors to the TLS model of Equation~\ref{eq:QiFit} and obtain $F \delta^0_{\mathit{TLS}}$, presented in figure~\ref{fig:resonators}(b) for all of the resonators. For the resonators on the \SI{500}{\nano\meter} film, $F \delta^0_{\mathit{TLS}}$ has settled at a lower level (\SI{5e-7}{} on average) in comparison with those made with the \SI{150}{\nano\meter} film (\SI{1e-6}{} on average), indicating a reduction in the loss due to the TLSs. As for the resonators made with the \SI{300}{\nano\meter} Al, $F \delta^0_{\mathit{TLS}}$ is in-between, with an average value of \SI{8e-7}{}.   

From the fit we also obtain $\delta_0$ in Equation~\ref{eq:QiFit}, see figure~\ref{fig:resonators_HP} in section \ref{sup:ResHP}. Interestingly, we observe a frequency-dependent increase in the non-TLS related loss, $\delta_0$, in the \SI{500}{\nano\meter} thick films. 
We did not observe this increased loss in our qubits, whose frequencies lie below \SI{5}{\giga\hertz}, where this loss seems to be less prominent. However, depending on its origin as well as its dependence on device geometry, this loss mechanism could become limiting for higher-frequency qubits fabricated on thicker films. We present a further investigation of this loss in section \ref{sup:ResHP}.

In the model for interacting TLSs, the parameter $\beta$ describes how sharply the TLSs saturate with increased power. 
In our data, $\beta$ is scattered, falling between 0.15 - 0.4 for most resonators (which is typically reported in studies \cite{burnett2017}), with no discernible trend across the devices.

\begin{figure}
    \centering
    \includegraphics[width=8.5cm]{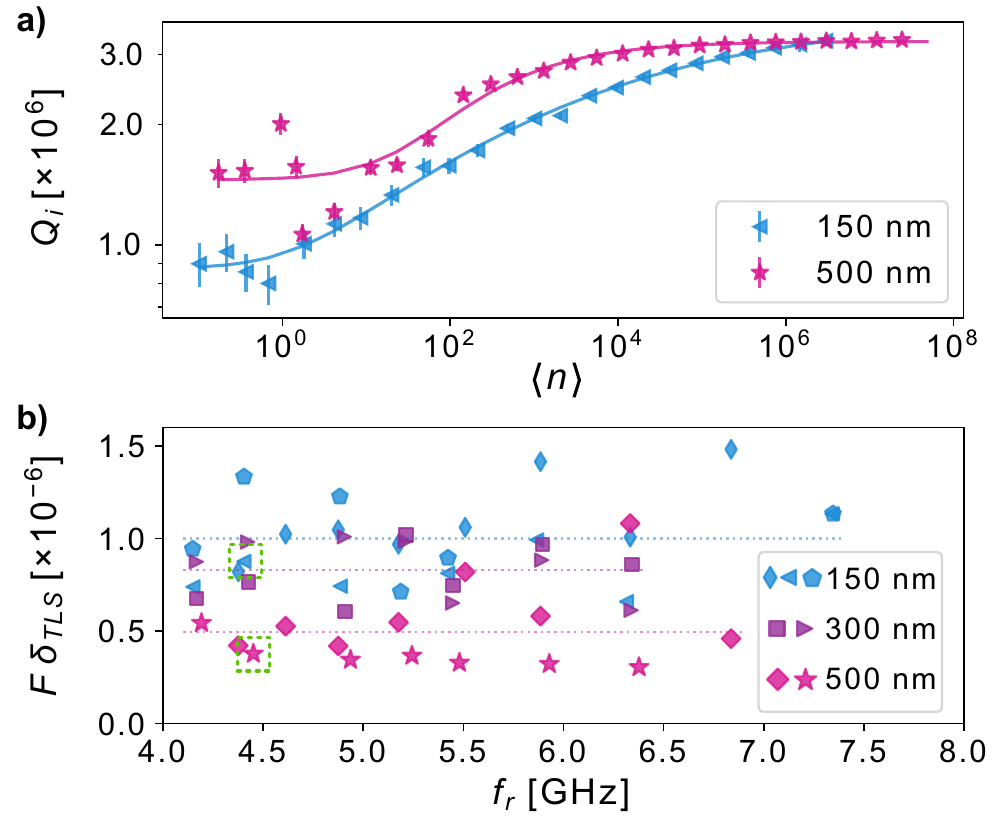}
    \caption{\label{fig:resonators} Study of energy loss in bare resonators. 
    \textbf{a)} Internal quality factor $Q_i$ as a function of average number of photons $\langle n \rangle$ fitted to the  TLS model in Equation~\ref{eq:QiFit}, for two resonators with a resonant frequency of \SI{4.45}{\giga\hertz}. \textbf{b)} TLS loss $F \delta_{\mathit{TLS}}$  extracted from the TLS model fits, as a function of frequency for resonators of three different film thicknesses. Data indicated by the same marker belong to resonators on a single chip. The dashed lines indicate the average value of $F \delta_{\mathit{TLS}}$ for a given film thickness.  For best visual comparability, the resonators showcased in \textbf{(a)} are chosen for their proximity in frequency and high-power $Q_i$, and are indicated in \textbf{(b)} with a dashed-line square.
    }
\end{figure}

\section{Materials analysis}{\label{sec:materials}

The presence of parasitic TLS defects is mainly attributed to amorphous oxides at the materials interfaces \cite{Martinis2005,lisenfeld2015}, although their microscopic origin is debated. We therefore perform an elemental analysis of representative samples, focusing on oxygen (O) residing near the material interfaces. We perform this analysis using time-of-flight secondary ion mass spectrometry (ToF-SIMS), a highly accurate method capable of simultaneous analysis of species with different masses \cite{Benninghoven}. To obtain a depth profile of the species of interest across all interfaces, we use an auxiliary sputter beam. This method is further detailed in section~\ref{sup:SIMS}.

The samples under analysis are pieces of a Si wafer coated with Al films of the three different thicknesses representative of our devices (150, 300, and \SI{500}{\nano\meter}). To discount wafer-to-wafer variation in substrate material properties, these three pieces originate from the same wafer. After the splitting of the wafer, the Si pieces underwent a cleaning and deposition procedure identical to that used during the qubit and resonator fabrication. 

The intensity of the detected signal identifying Si, Al, and O across the thickness of the sample is plotted in figure~\ref{fig:SIMS}, normalized to the total ion count. Each trace starts at the metal-air (MA) interface, then continues through the metal, and ends inside the substrate. 

In figure~\ref{fig:SIMS}(a), we detect Si at the MA interface,  which is likely coming from ambient air}. Sputtering away more material of the stack, the intensity of Si is low inside the metal, then rises abruptly at the SM interface, and saturates inside the Si substrate. In figure~\ref{fig:SIMS}(b), the intensity of the Al signal is initially high at the MA interface and inside the Al film, then plummets to the detection limit once we reach the substrate.

\begin{figure}
    \centering
    \includegraphics[width=8.5cm]{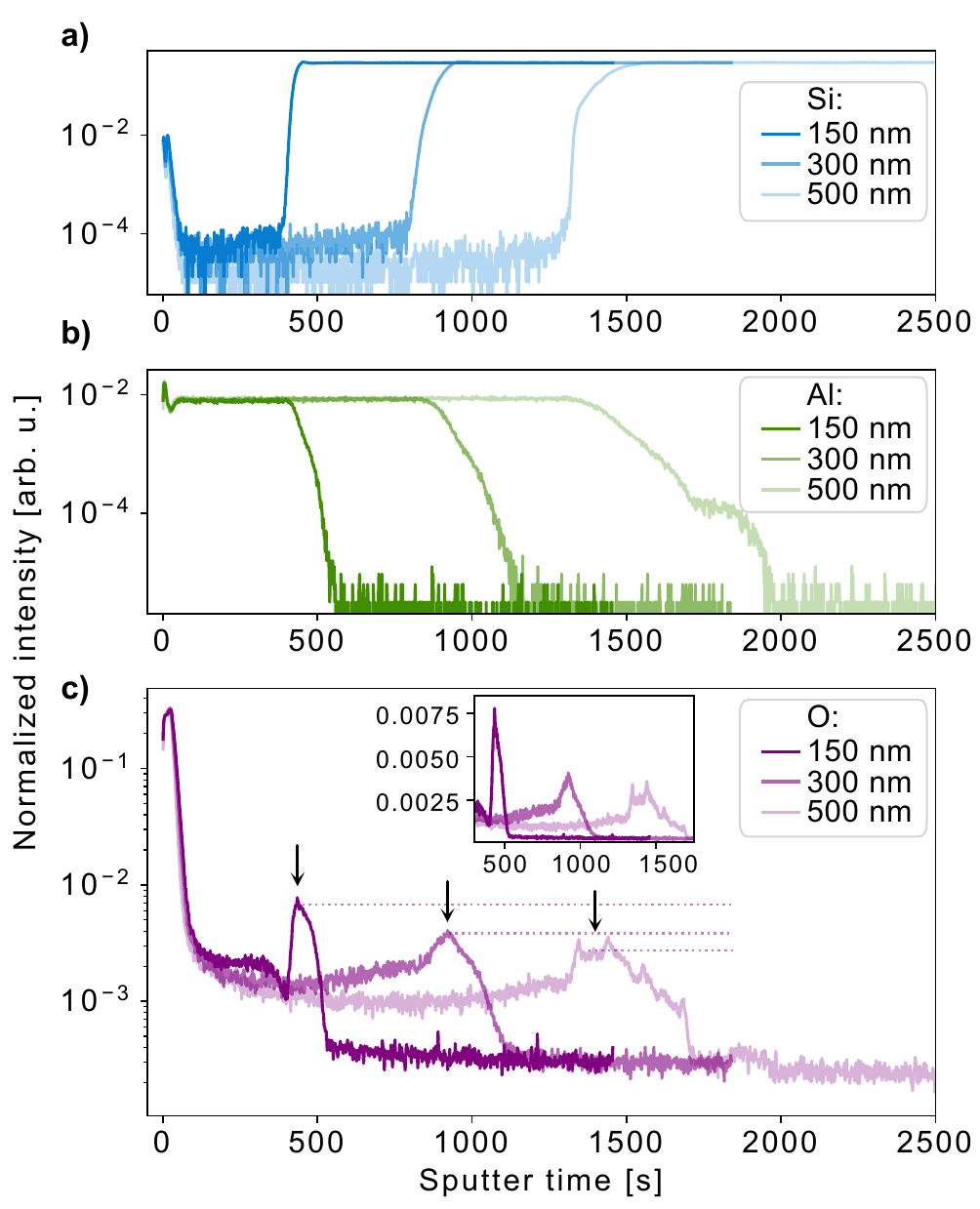}
    \caption{\label{fig:SIMS} 
    Elemental analysis of the Al/Si material stack by ToF-SIMS. Depth profiles of \textbf{a)} Si, \textbf{b)} Al and \textbf{c)} O. Each plot contains data on three samples with varying Al thickness. The sharp increase in the intensity of Si in \textbf{(a)}, and the arrows in \textbf{(c)}, mark the SM interface of each sample. The inset in \textbf{(c)} shows the O peaks at the SM interface on a linear scale.}
\end{figure}

In figure~\ref{fig:SIMS}(c), the intensity of the O signal starts high at the aluminum oxide-passivated MA surface, then stabilizes at a lower level inside the film. Sputtering further, there is a peak in the O intensity at the SM interface, after which the trace levels out inside the Si substrate.

The diffusion constant of oxygen in aluminum is low at room temperature \cite{interstitialO}; however, diffusion along grain boundaries is substantially faster \cite{Jaseliunaite2020}. Oxygen can exist in aluminum along the grain boundaries, as the evaporated Al grows in high vacuum  with a tendency to follow a columnar growth structure \cite{Adamik1998}. This observation emphasizes that lossy oxides exist not only at the direct interfaces, but also within the grainy metal films, contributing to dielectric loss when in the vicinity of the interfaces.

We suggest that the peak at the SM originates from both the continuation of oxide along the grains, as well as from an interfacial oxide. The origin of this interfacial oxide could be a result of incomplete silicon oxide removal, or oxide regrowth after removal during transfer to the evaporator or while inside, resulting from residual oxygen content of the evaporation chamber. In section~\ref{sup:SIMS} we show that the oxygen is mainly present in the form of aluminum oxide, which is consistent with the thermodynamically favorable reduction of silicon oxides into aluminum oxides \cite{SiOxReduction1,SiOxReduction2}.

Comparing the three films in figure~\ref{fig:SIMS}(c), we find that the oxygen level inside the film decreases in the thicker films. The peak intensity is also lower in the thicker films, as indicated with the dashed lines.

While it appears in the data as if the Si/Al interface were broadening with increasing film thickness, this is unlikely to be a real feature of the samples. Instead, this broadening is a known artifact caused by surface roughening and atomic mixing during the sputter-assisted depth profiling \cite{Yan2019,Gorbenko2016}. Although disregarding the width of the peaks in similar scenarios is not unusual, we compare both the peak intensity and the integrated intensity for O across the entire Al film. We find a similar trend in both cases, that the intensity of the oxygen signal is strongest for the thinnest film. At the SM interface, the O intensity peak level scales with the ratio of 5:3:2 for \SI{150}{\nano\meter}, \SI{300}{\nano\meter}, and \SI{500}{\nano\meter} films, respectively. The integrated intensity for O across the entire Al film, including the MA and SM interfaces, divided by film thickness scales with a ratio of 6:3:2.

Increasing a film's thickness has a direct effect on the resulting grain size \cite{Chaverri1991,Nik2016}, which we confirm using transmission electron microscopy (TEM). In figure~\ref{fig:TEM}, we observe long vertical grain boundaries which stretch uninterrupted all the way from the top surface to the interface with the Si substrate. The grain size increases with film thickness, which strongly impacts the morphology of the material interfaces. A larger grain size also provides fewer grain boundaries to trap residual oxygen during deposition, as well as to act as diffusion channels upon contact with ambient atmosphere. While the electric field well inside the superconducting film is low, and therefore grain boundaries inside the film are unlikely to lead to additional loss, fewer grain boundaries acting as hosts of TLS in the vicinity of the SM interface can decrease the dielectric loss of this interface.}
Since the contribution of the SM interface to the overall loss is substantial (see section \ref{sec:participation}), a small change in the concentration of potential TLS defects has a significant impact on the total loss in the circuit. 

At the MA interface, the oxygen contribution from the surface oxide is considerably greater than the oxide along the grain boundaries. As such, we do not observe a tangible effect in the reduction of oxygen presence between the samples with different densities of grain boundaries. The measurement traces identifying oxygen species at the surface of the three different film thicknesses are indistinguishable. 
Altering the thickness of a thin film can affect the MA interface as well, by way of changing the roughness of the film's top surface. We perform an atomic force microscopy (AFM) to evaluate the surface roughness of a \SI{150}{\nano\meter} and a \SI{500}{\nano\meter} film, obtaining root mean square roughness $R_q$ values of \SI{5.9}{\nano\meter} and \SI{6.3}{\nano\meter}, respectively, over a $5 \times \SI{5}{\micro\meter\squared}$ area (figure~\ref{fig:AFM}). This negligible difference, in conjunction with no observed change in MA oxide in the SIMS measurement, further discounts this interface from explaining the coherence measurement data.

\begin{figure}
    \centering
    \includegraphics[width=8.5cm]{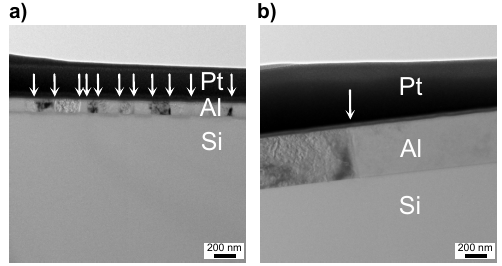}
    \caption{\label{fig:TEM} 
     Investigation of the grain boundaries. TEM images showing the grains of \textbf{a)} \SI{150}{\nano\meter} and \textbf{b)} \SI{500}{\nano\meter} thick films. The grain size increases with film thickness, resulting in fewer grain boundaries. The arrows show where the grain boundaries are. The platinum (Pt) layer is added during sample preparation for TEM.}
\end{figure}

\section{Participation ratio simulations}{\label{sec:participation} 

\begin{figure}
    \centering
    \includegraphics[width=8.5cm]{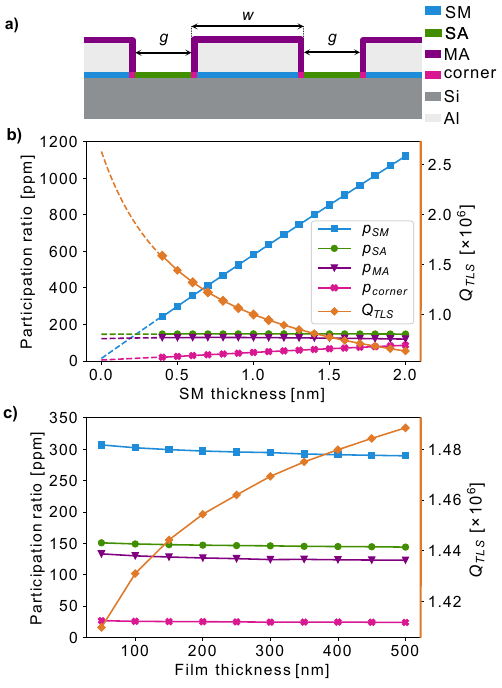}
    \caption{\label{fig:simulation} 
     CPW geometry and participation ratio simulations. \textbf{a)} CPW cross section indicating the modeled interfaces. \textbf{b)} 2D simulation results of a CPW resonator showing the variations in the participation ratio at different interfaces as a function of the dielectric thickness at the substrate-metal (SM) interface. The thickness of the metallic film is kept fixed at \SI{150}{\nano\meter}. \textbf{c)} Variation in the participation ratios as a function of the thickness of the metallic film of the CPW resonator. In both \textbf{(b)} and \textbf{(c)}, the corresponding theoretical internal quality factor due to TLS loss is shown in orange (right axis). Note that although the majority share of the electric field is stored in the substrate (91-92\%) and in air (8-9\%), their corresponding participation ratios are not shown in the figure; these two media are not as lossy as the other interfaces (table~\ref{tab:ParticipationRatio}), and they do not hold a major part of the total loss. }
\end{figure}

In this section, we present energy participation ratio simulations of a 2D CPW geometry, showing that the reduced loss at the SM interface is the most likely reason for the improved quality factors of the devices made with thicker aluminum films. We also investigate if and when the loss due to other interfaces becomes dominant.

\begin{table}
\begin{ruledtabular}
\caption{\small{\label{tab:ParticipationRatio} Parameters used for participation ratio simulations in figure~\ref{fig:simulation}. For simplicity, the air environment (cryogenic vacuum) is assumed lossless and the values of $\text{tan}\,\delta$ for SM, SA and MA interfaces are considered equal. $\text{tan}\,\delta = \text{10}^\text{-3}$ stands at the lower bound of the limits reported by Wang et al. \cite{Wang2015}. The CPW dimensions are $w = \SI{20}{\micro\meter}$ and $g = \SI{10}{\micro\meter}$}.}
\renewcommand{\arraystretch}{1.1}
\centering\begin{tabular}{l l r r r }
 \\[-1.8ex]
   Interface & $\text{tan}\,\delta$ & $\epsilon_r $ & Thickness & Thickness \\
    &  &  & Fig~\ref{fig:simulation}\textbf{(b)} & Fig~\ref{fig:simulation}\textbf{(c)} \\[3pt]
   \hline \\[-1.8ex]
   Al & \, - & - & \SI{150}{\nano\meter} & 50 - \SI{500}{\nano\meter} \\
   Air &  \, 0 & 1.0 & \SI{2}{\milli\meter} & \SI{2}{\milli\meter} \\
   Si &  $\text{10}^{-7}$ & 11.7 & \SI{280}{\micro\meter} & \SI{280}{\micro\meter} \\
   MA &  $\text{10}^{-3}$ & 7.0 & \SI{5}{\nano\meter} & \SI{5}{\nano\meter} \\
   SA &  $\text{10}^{-3}$ & 4.0 & \SI{2}{\nano\meter} & \SI{2}{\nano\meter} \\
   SM &  $\text{10}^{-3}$ & 4.0 & 0.4 - \SI{2}{\nano\meter} & \SI{0.5}{\nano\meter} \\
   corner &  $\text{10}^{-3}$ & 4.0 &  & \\[2pt]

\end{tabular}
\end{ruledtabular}

\end{table}

The cross section of the simulated device is illustrated in figure~\ref{fig:simulation}(a), with the center conductor width {$w = \SI{20}{\micro\meter}$ and  the gap to ground {$g = \SI{10}{\micro\meter}$, which corresponds to a $\sim\SI{50}{\ohm}$ impedance. Using the material parameters in table~\ref{tab:ParticipationRatio} \cite{Martinis2014,Wang2015}, we first sweep the thickness of the dielectric at the SM interface and obtain the participation ratios presented in figure~\ref{fig:simulation}(b). The data points show the simulation results from \SI{2}{\nano\meter} down to \SI{0.4}{\nano\meter}, while the lines represent polynomial fits to estimate the participation ratios for a thinner dielectric.  
The total loss due to TLSs, i.e., $1/Q_\mathit{TLS}$ -- with $Q_\mathit{TLS}$ presented in figure~\ref{fig:simulation}(b) as well -- scales with the participation ratio weighted by the loss tangent of the material

\begin{equation}\label{eq:1_Q-losstangents}
    1/Q_\mathit{TLS} = \sum_i p_i \tan \delta_i \,,
\end{equation}

\noindent where $p_i$ and $\text{tan}\,\delta_i$ denote the participation ratio and the loss tangent at interface $i$, respectively ($i$: Si, MA, SA, SM and corner). The various interfaces are illustrated in figure~\ref{fig:simulation}(a).

The energy participation ratio at the MA and SA interfaces changes negligibly by varying the dielectric thickness at the SM. However, at the SM interface (and the corresponding corners), there is a linear dependence of the field concentration on the SM dielectric thickness. Moreover, compared to the MA and SA interfaces, a considerable fraction of the electric field resides at the SM interface. With the removal of the silicon native oxide prior to metal deposition, the \textit{effective} dielectric thickness at the SM interface is estimated to be about \SI{0.5}{\nano\meter} \cite{Kosen2022}. The resulting participation ratio, and the corresponding loss, assuming equal loss tangents, is about twice those at the MA and SA interfaces. Reducing the SM dielectric to about \SI{0.25}{\nano\meter} decreases the corresponding SM loss, yielding an almost equal value of the participation ratio by the three interfaces. Below this level, the loss is dominated by the MA and SA interfaces. 

We ran a second round of simulations to explore the evolution of participation ratios if the thickness of the metallic layer is varied. The presented results in figure~\ref{fig:simulation}(c) show only a slight variation in the participation ratios, as well as in the quality factor of the resonator. From the simulations above and the materials analysis in section~\ref{sec:materials}, we conclude that the better performance of the devices with the thicker metallic layers is due to the reduction of the dielectric loss at the SM interface, by way of reducing the number of grain boundaries.


\section{Conclusion}
We observe a significant improvement in the energy relaxation time of transmon qubits on silicon substrates when fabricated using aluminum films thicker than \SI{300}{\nano\meter}, whereas the previous standard in our laboratory, and generally in the field, has been thinner. This observation is based on measurements of 36 qubits and 50 resonators.
We demonstrate transmon qubits with average relaxation times exceeding \SI{200}{\micro\second} on \SI{500}{\nano\meter} films, with the best qubit showing a time-averaged $T_1 =\,$\SI{270}{\micro\second}, corresponding to $Q=\text{5.1}\times \text{10}^\text{6}$.

Despite removing the native oxide of the silicon substrate prior to Al deposition, we detect a presence of oxygen and aluminum oxide at the Al--Si interface, which, due to the relatively strong electric field at this interface, contributes to dielectric loss. Material-depth characterization by ToF-SIMS reveals a weakening intensity of the oxygen-signal peak at the substrate-metal interface with increasing film thickness from 150 to 300 to \SI{500}{\nano\meter}. 

We attribute the lower intensity of oxygen at the SM interface of thicker films, hence their lower dielectric loss, to the increased grain size in the thicker film, which we confirm in a TEM analysis. 
This consequently reduces the prevalence of oxidized grain boundaries in the vicinity of the interface, leading to lower loss.

From a study of loss in CPW resonators, we conclude that the contribution from TLS defects decreases for the thicker films and becomes comparable to loss due to other mechanisms.
Whether growing even thicker films---or indeed using epitaxial films---can deliver qubits of superior performance demands further investigations. Our experiments and simulations point towards two obstacles: Firstly, a frequency-dependent loss appears for the thickest film of \SI{500}{nm}. 
Secondly, with the reduced dielectric thickness at the SM interface (below about \SI{0.2}{\nano\meter}), the corresponding interfacial loss falls below that of the other interfaces, with a lower impact on the total loss.

\begin{acknowledgments}
We are grateful for discussions with Liangyu Chen, Simone Gasparinetti, Stefan Gustafsson, Lars Jönsson, Mikael Kervinen, Zeinab Khosravizadeh, Sandoko Kosen, Sergey Kubatkin, David Niepce, Andreas Nylander, Eva Olsson, Marco Scigliuzzo, Daryoush Shiri, Giovanna Tancredi and Lunjie Zeng. 

The device fabrication was performed at Myfab Chalmers. The ToF-SIMS measurements were performed at the Chemical Imaging infrastructure at Chalmers University of Technology. The TEM was performed at the Chalmers Materials Analysis Laboratory.

This work was funded by the Knut and Alice Wallenberg (KAW) Foundation through the Wallenberg Center for Quantum Technology
(WACQT), and by the EU Flagship on Quantum Technology HORIZON-CL4-2022-QUANTUM-01-SGA project 101113946 OpenSuperQPlus100.
\end{acknowledgments}

\appendix

\section{\label{sup:fab}Device design and fabrication}

In figure~\ref{fig:qubits} we show a representative qubit design. It consists of a Josephson junction shunted by a large cross-type capacitor, which is coupled capacitively to the open end of a quarter-wavelength ($\lambda$/4) CPW resonator. This resonator is in turn inductively coupled  via the shorted end to the input/output transmission line. 

For readout resonators, both the center conductor width $w$ and the gap $g$ to the ground are \SI{12}{\micro\meter}, while for the qubit capacitor {$w = g = \SI{24}{\micro\meter}$. The device structures are surrounded by an array of flux-trapping holes \cite{Chiaro2016}. These are squares with a \SI{2}{\micro\meter} long side, and a \SI{10}{\micro\meter} pitch.
The bare resonators used for energy loss quantification are also of the $\lambda$/4 CPW type, with $w = 2\,g = \SI{20}{\micro\meter}$, coupled capacitively to the transmission line. 

\begin{figure*}
    \includegraphics[width=14cm]{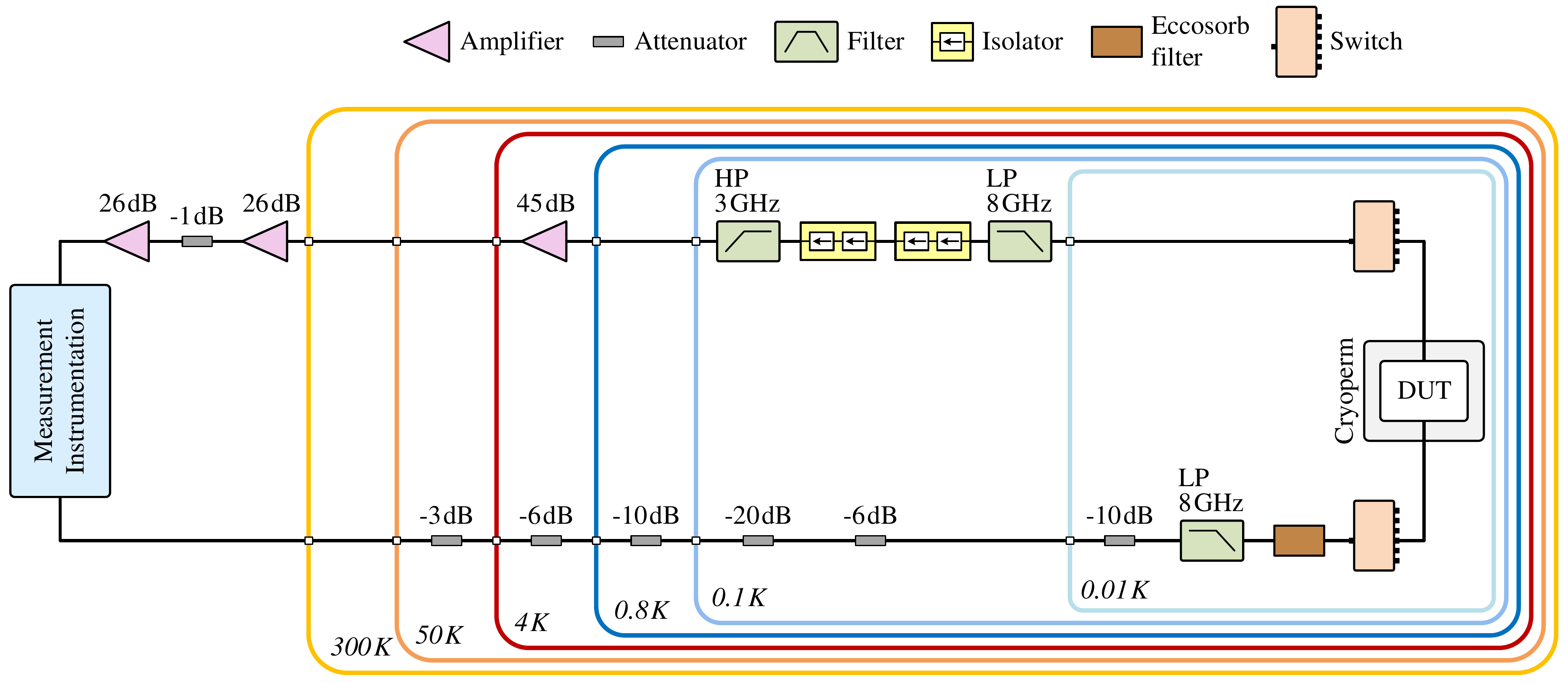}
    \caption{\label{fig:sup_setup} Cryogenic measurement setup. }
\end{figure*}

The devices are fabricated on high-resistivity ($\rho \geq \SI{10}{\kilo\ohm\centi\meter}$) intrinsic silicon substrates.
The substrates are stripped of their native oxide in a 2\% aqueous solution of HF and subsequently rinsed with deionized water to leave the Si surface terminated with a hydrogen monolayer \cite{HFdip}.

Immediately after rinsing, we load the Si wafer into the load-lock of a Plassys MEB 550s evaporator, where it is heated to \SI{300}{\celsius} for \SI{10}{\minute} and then left to cool down to room temperature. This is done to desorb impurities and moisture from the surface of the wafer, and to allow the evaporator to reach a satisfactory vacuum level of $\sim$\SI{3e-8}{\milli\bar} in the evaporation chamber.
We deposit the aluminum films at the rate of \SI{1}{\nano\meter\per\second}, and then oxidize the top surface \textit{in situ} in the load-lock without breaking the vacuum, in a static oxidation step.

We then define all circuitry barring the Josephson junctions (JJs) in an optical lithography step. This includes the transmission line, the coplanar waveguide (CPW) resonators, qubit capacitors, and flux trapping holes. We use a resist stack of an e-beam resist, PMMA\,A2, and an optically sensitive resist such as S1805. The PMMA layer underneath the optical resist serves to protect the underlying Al from damage induced by the TMAH present in the photo-developer (MF319), which enables re-patterning. After successful patterning, the PMMA is ashed away in oxygen plasma. The pattern defined in the resist stack is then transferred into the Al layer via wet etching in a mixture of phosphoric, nitric, and acetic acids (aluminum etchant type A).

For qubit devices, JJs are fabricated using the patch-integrated cross-type technique adapted from \cite{PICT}. The JJ electrodes pattern is transferred to the MMA\,EL12 + PMMA\,A6 resist stack using e-beam lithography (EBL), followed by deposition of the electrodes' metal in the Plassys evaporator using shadow evaporation and planetary turn. The thickness of the bottom electrode is \SI{50}{\nano\meter} and the top electrode is \SI{110}{\nano\meter} thick. A liftoff in Remover\,1165 completes the fabrication. 

\section{\label{sup:setup}Measurement setup}

We perform all of the microwave measurements at $\sim$\SI{10}{\milli\kelvin}. The devices under test are placed in a light-tight copper sample box, mounted on a copper tail attached to the mixing chamber stage of a Bluefors LD250 dilution refrigerator. The copper tail is enclosed by a copper can coated with a layer of Stycast mixed with carbon and silicon carbide on the inside, which is in turn enclosed by a Cryoperm magnetic shield. 

Figure~\ref{fig:sup_setup} shows the cryogenic measurement setup with all components including attenuators, filters, microwave switches, isolators and amplifiers. Apart from the attenuators marked in the diagram, the input line itself attenuates by an additional \SI{11}{\decibel}, as determined from an ac-Stark shift measurement \cite{TWPA_macklin,bruno2015}. 

As for the room temperature instrumentation, all resonator spectroscopy measurements necessary for the $Q$ factor extraction are performed using a R\&S ZNB8 vector network analyzer (VNA). When necessary, additional attenuation up to \SI{-20}{\decibel} is placed at the VNA output, in order to reach powers equivalent to a single photon circulating in the resonators.

The qubits are characterized using two alternative measurement setups. One setup consists of a multi-frequency lock-in amplifier platform (Intermodulation Products Presto-16), where both the pulses necessary to control the qubit and read out its resonator are synthesized directly on an FPGA. The measured signal is also digitized on this FPGA. The alternative qubit measurement setup consists of high frequency signal generators (R\&S SGS100a) and arbitrary waveform generators (Keysight PXIe M3202A 1GS/s AWG), whose outputs are upconverted with the help of local oscillators to generate the desired high frequency pulses. The detected signal is digitized using a Keysight PXIe M3102A 500 MS/s digitizer. Both setups have been verified to yield identical coherence results.

\section{\label{sup:QubitsTable}Summary of qubit measurements}

We measured 38 qubits fabricated on 8 separate wafers. The key parameters such as the aluminum film thickness, qubit frequency $\omega_q/2\pi $, readout resonator frequency $\omega_r/2\pi $, the qubit-resonator detuning $\Delta$, the arithmetic mean values (plus/minus one standard deviation) for the relaxation time $T_1$ and the spin echo decoherence time $T_2^\text{\,echo}$, the qubit quality factor $Q$, and the calculated Purcell-decay time $T_p$, are summarized in table~\ref{tab:qubits}.

Two qubits with possibly inaccurately obtained parameters are not included in the analysis in figure~\ref{fig:qubits}. These are \textit{Q22} ($T_2^{\text{\,echo}} > 2\,T_1$) and \textit{Q38} ($T_1 > T_p$). This could be due to parameter fluctuations during the measurement time span. 

For \textit{Q38}, the measured $T_1$ exceeded the $T_p$ inferred from the device parameters. This is possibly due to the asymmetry of the resonance dip caused by an impedance mismatch. This gives the qubit some protection from decay, which is not accounted for in the model used for the Purcell decay calculations. 


\begin{table*}
\caption{\small{\label{tab:qubits} Summary of qubit measurements. Qubits with identical markers are from the same wafer. Qubits without a marker are excluded from the analysis in figure~\ref{fig:qubits}. }}

\centering\begin{tabular}{c >{\centering}m{0.7cm} >{\centering}m{1.6cm} >{\centering}m{1.3cm} >{\centering}m{1.3cm} >{\centering}m{1.3cm} >{\centering}m{1.8cm} >{\centering}m{1.8cm} >{\centering}m{1.3cm} c}
 
\hline\hline\\[-1.ex]
   &  & Thickness & $\omega_q/2\pi $ & $\omega_r/2\pi $ & $\Delta$ & $T_1 $ &  $T_2^\text{\,echo} $ & $T_p$ & $Q $ \\
   &  & $ \text{(nm)}$ & $ \text{(GHz)}$ & $ \text{(GHz)}$ & $ \text{(GHz)}$ & $ \text{(}\mu\text{s)}$ & $ \text{(}\mu\text{s)}$ & $ \text{(}\mu\text{s)}$ & $(\times \textbf{10}^6)$ \\[6pt]
   
   \hline 
   &  &  &  &  &  &  &  &  &  \\
   \tiny{$\times$}& Q1 & 150 & 4.711 & 6.035 & 1.324 & 51 $\pm$ 9  & 79 $\pm$ 17  &64  & 1.5 \\ 
   \tiny{$\times$}& Q2 & 150 & 4.662 & 6.173 & 1.511 & 58 $\pm$ 14  & 100 $\pm$ 25  &155 & 1.7 \\
   \tiny{$\times$}& Q3 & 150 & 4.923 & 6.317 & 1.394 & 39 $\pm$ 6  & 66 $\pm$ 11  &95  & 1.2 \\
   \tiny{$\times$}& Q4 & 150 & 4.721 & 6.039 & 1.318 & 70 $\pm$ 11 & 110 $\pm$ 30 &155 & 2.1 \\
   \tiny{$\times$}& Q5 & 150 & 4.634 & 6.175 & 1.541 & 62 $\pm$ 18  & 101 $\pm$ 37  &151 & 1.8 \\
   \tiny{$\times$}& Q6 & 150 & 4.449 & 6.320 & 1.871 & 98 $\pm$ 27  & 135 $\pm$ 46 &118 & 2.7 \\
   $\smallblacktriangledown$& Q7 & 150 & 3.650 & 6.036 & 2.386 & 124 $\pm$ 33 & - & 255 & 2.8\\
   $\smblkcircle$& Q8 & 150 & 4.689 & 6.401 & 1.712 & 62 $\pm$ 13  & - &223 & 1.8 \\
   $\smblkcircle$& Q9 & 150 & 4.388 & 6.036 & 1.648 & 82 $\pm$ 17  & - &370 & 2.3 \\
   $\smblkcircle$& Q10& 150 & 4.251 & 6.102 & 1.851 & 92 $\pm$ 28  & - &436 & 2.5 \\
   $\smblkcircle$& Q11& 150 & 4.126 & 6.259 & 2.133 & 89 $\pm$ 25  & - &192 & 2.3 \\
   $\smblkcircle$& Q12& 150 & 3.939 & 6.097 & 2.158 & 99 $\pm$ 28  & - &384 & 2.5 \\
   $\smblkcircle$& Q13& 150 & 3.742 & 6.254 & 2.512 & 81 $\pm$ 30 & - &512 & 1.9 \\
   $\smallblacksquare$& Q14& 300 & 4.510 & 6.075 & 1.564 & 92 $\pm$ 28  & 106 $\pm$ 26 & 127 & 2.6 \\
   $\smallblacksquare$& Q15& 300 & 4.221 & 6.212 & 1.991 & 113 $\pm$ 23  & 135 $\pm$ 22 & 421 & 3.0 \\
   $\smallblacksquare$& Q16& 300 & 3.883 & 6.355 & 2.473 & 160 $\pm$ 39  & 175 $\pm$ 30 & 545 & 3.9 \\
   $\smallblacksquare$& Q17& 300 & 4.027 & 6.070 & 2.043 & 133 $\pm$ 18  & 101 $\pm$ 14 & 247 & 3.4 \\
   $\smallblacksquare$& Q18& 300 & 3.692 & 6.350 & 2.658 & 157 $\pm$ 40  & 228 $\pm$ 71 & 716 & 3.6 \\
   $ \smallblacktriangleup$& Q19& 300 & 4.899 & 6.072 & 1.173 & 67 $\pm$ 15  & 61 $\pm$ 15 & 94 & 2.1 \\
   $ \smallblacktriangleup$& Q20& 300 & 4.490 & 6.209 & 1.719 & 133 $\pm$ 25  & 116 $\pm$ 21 & 202 & 3.8 \\
   $ \smallblacktriangleup$& Q21& 300 & 4.283 & 6.352 & 2.069 & 106 $\pm$ 17  & 146 $\pm$ 24 & 245 & 2.9 \\
   & Q22& 500 & 4.729 & 6.113 & 1.384 & 64 $\pm$ 18  & 158 $\pm$ 58 & 108 & 1.9 \\
   $\blackdiamond$& Q23& 500 & 4.421 & 6.251 & 1.830 & 135 $\pm$ 33  & 159 $\pm$ 56 &211 & 3.8 \\
   $\blackdiamond$& Q24& 500 & 4.176 & 6.396 & 2.220 & 156 $\pm$ 43 & 165 $\pm$ 54 &285 & 4.1 \\
   $\star$& Q25& 500 & 2.873 & 6.250 & 3.377 & 248 $\pm$ 109  & 216 $\pm$ 90 &1678& 4.5 \\
   $\star$& Q26& 500 & 3.164 & 6.393 & 3.229 & 159 $\pm$ 44  &  9 $\pm$ 1 &1360 & 3.2 \\
   $\star$& Q27& 500 & 3.016 & 6.386 & 3.371 & 270 $\pm$ 83  & 307 $\pm$ 114 &1141 & 5.1 \\
   $\star$& Q28& 500 & 3.206 & 6.108 & 2.902 & 152 $\pm$ 31  & 196 $\pm$ 33 &766 & 3.1 \\
   $\star$& Q29& 500 & 3.490 & 6.246 & 2.756 & 168 $\pm$ 31 & 248 $\pm$ 48 &739 & 3.7 \\
   $\star$& Q30& 500 & 2.776 & 6.244 & 3.468 & 196 $\pm$ 39 & 177 $\pm$ 39 & 1909 & 3.4 \\
   $\star$& Q31& 500 & 3.895 & 6.390 & 2.495 & 155 $\pm$ 52  & 252 $\pm$ 80 & 505 & 3.8 \\
   \scriptsize{$\plus$}& Q32& 500 & 4.641 & 6.304 & 1.663 & 60 $\pm$ 19  & 90 $\pm$ 30 & 73 & 1.7 \\
   \scriptsize{$\plus$}& Q33& 500 & 4.810 & 6.577 & 1.767 & 79 $\pm$ 14  & 99 $\pm$ 19 & 120 & 2.4 \\
   \scriptsize{$\plus$}& Q34& 500 & 4.672 & 6.440 & 1.768 & 84 $\pm$ 22  & 143 $\pm$ 50 & 236& 2.5 \\
   \scriptsize{$\plus$}& Q35& 500 & 4.798 & 6.722 & 1.924 & 83 $\pm$ 21  & 110 $\pm$ 27 & 140& 2.5 \\
   \scriptsize{$\plus$}& Q36& 500 & 4.750 & 6.303 & 1.553 & 46 $\pm$ 13 & 75 $\pm$ 23 & 85& 1.4 \\
   \scriptsize{$\plus$}& Q37& 500 & 4.830 & 6.441 & 1.611 & 112 $\pm$ 30  & 151 $\pm$ 39 & 138& 3.4 \\
   & Q38& 500 & 5.025 & 6.578 & 1.553 & 77 $\pm$ 13  & 122 $\pm$ 26 & 61 &2.4 \\[6pt]
   
\hline\hline
  \end{tabular}

\end{table*}

\section{\label{sup:ResHP}High-power loss of thicker films}

In section \ref{sec:resonators} we observe that TLS-related loss, which shows itself in the suppressed $Q_i$ of CPW resonators when the average circulating photon number $\langle n \rangle$ is low, is mitigated in thicker aluminum films. However, we also observe that the energy loss due to other mechanisms, which imposes an upper limit for $Q_i$ at high $\langle n \rangle$, increases for the CPWs fabricated on \SI{500}{\nano\meter} thick films. Similarly to TLS loss, we quantify these other losses by fitting the experimental data to Equation~\ref{eq:QiFit}. The extracted parameter $\delta_0$ representing non-TLS-related losses is shown in figure~\ref{fig:resonators_HP}. At high frequencies, this loss becomes a limiting factor of the resonators made on thicker films.

\begin{figure}
    \centering
    \includegraphics[width=8.5cm]{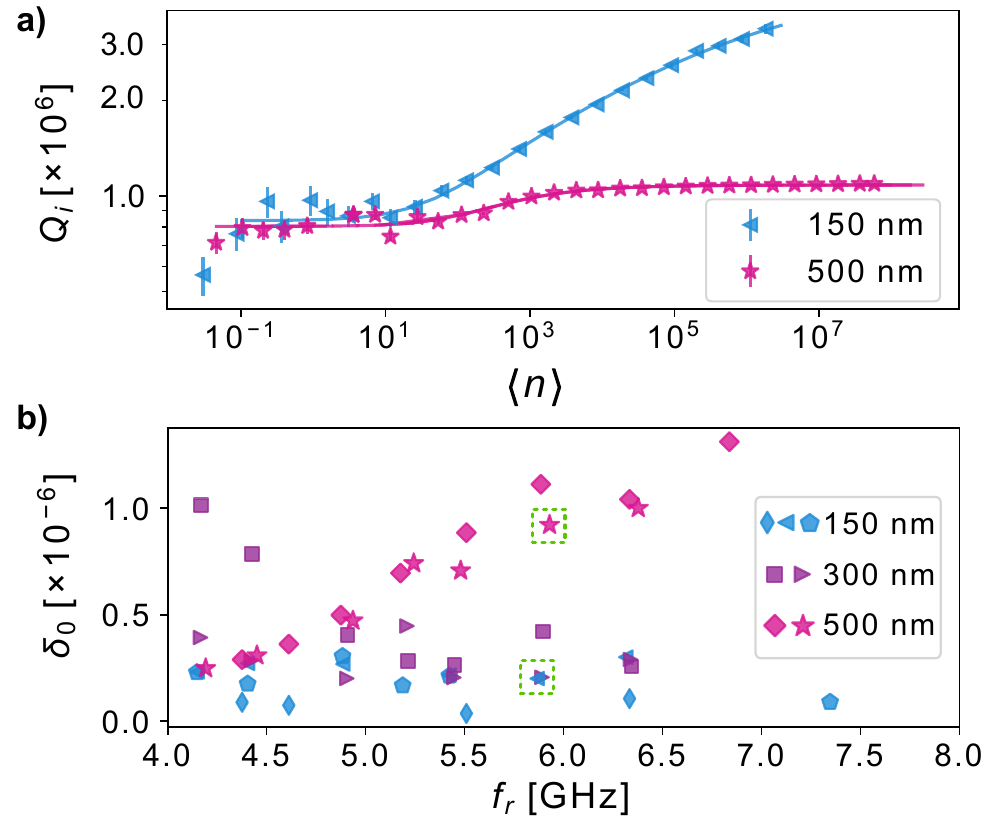}
    \caption{\label{fig:resonators_HP} Study of energy loss in bare resonators. 
    \textbf{a)} Internal quality factor $Q_i$ as a function of average number of photons $\langle n \rangle$ fitted to the TLS model in Equation~\ref{eq:QiFit}, for two resonators with a resonant frequency of \SI{5.9}{\giga\hertz}. \textbf{b)} The loss due to other sources $\delta_0$ extracted from the TLS model fits, as a function of frequency for resonators of three different film thicknesses. Data indicated by same marker belong to resonators on a single chip. For best visual comparability, the resonators showcased in \textbf{(a)} are chosen for their proximity in frequency and low-power $Q_i$, and are indicated in \textbf{(b)} with a dashed-line square.}
\end{figure}

\begin{figure}
    \centering
    \includegraphics[width=8.5cm]{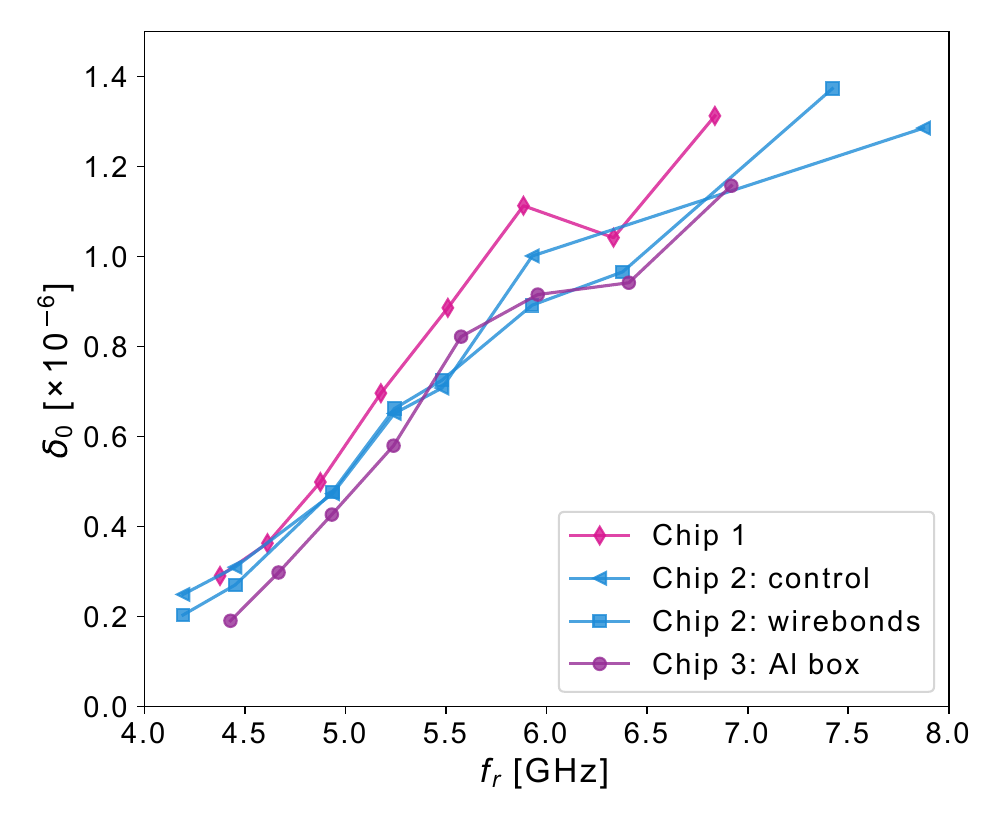}
    \caption{\label{fig:Resonators_HP_boxes} 
    High-power loss of resonators made on \SI{500}{\nano \meter} thick films. \textit{Chip\,1} and \textit{Chip\,2} are placed (each) in a copper sample box, while \textit{Chip\,3} is in an aluminum box.} 
\end{figure}

To find the source of the frequency-dependent loss, $\delta_0$ in Equation~\ref{eq:QiFit}, we measure resonators fabricated on the thick, \SI{500}{\nano\meter} Al film, under different scenarios. The first two chips labeled with \textit{Chip\,1} and \textit{Chip\,2} in figure~\ref{fig:Resonators_HP_boxes} are the control chips whose results are presented in figure~\ref{fig:resonators_HP}, as well. After the measurement, we wire-bonded across the main transmission line and the resonators on \textit{Chip\,2} in order to suppress parasitic slotline modes \cite{Abuwasib2013, Chen2014}. In addition, we placed another chip in an aluminum sample box instead of the usual copper box (\textit{Chip\,3}). We expected this to change the coupling between the resonators and the box mode, resulting in a different radiation into the box mode. We did not observe any noticeable difference in the $\delta_0$ obtained, from which we conclude that, at the moment, neither of these two (the slotline modes or radiation into the box mode) are limiting the coherence of our devices.

\section{\label{sup:SIMS} Additional materials analysis}

In ToF-SIMS, a primary focused ion beam is used to sputter elements away from the sample surface. The secondary ions ejected during sputtering are accelerated towards a mass spectrometer, where their time-of-flight and mass/charge ratios are analyzed. 

In our measurements we are interested in a depth profile of the various elements across the Si/Al material stack. Therefore, we use an auxiliary 3 keV Cs$^+$ sputter beam with a raster size of  \SI{300}{\micro \meter} $\times$ \SI{300}{\micro\meter} in conjunction with the primary 25 keV Bi$^{3+}$} ion beam used for the secondary ion detection. The raster area of the primary Bi$^{3+}$ ion beam is \SI{100}{\micro\meter} $\times$ \SI{100}{\micro\meter}. The Cs dose density at 2500 sputter seconds was \SI{6e17}{} ions/cm$^2$. To reduce the signal from residual gases in the chamber in the depth profile, t}he analysis is performed in high vacuum in the range of 2.6 -- \SI{3.0e-9}{\milli\bar}. 

We study the distribution of the detected oxygen between silicon and aluminum oxides at the Si/Al interface in figure~\ref{fig:SIMS_supp}. The presence of silicon oxides is very low, barely above the detection limit. On the other hand, the presence of aluminum oxide is substantial.

In an Si/\ce{SiO2}/Al interface, it is thermodynamically favourable for Al to reduce \ce{SiO2} and form \ce{Al2O3} instead \cite{Bauer1980,Zeng2013}:
\begin{equation}
    \ce{4 Al + 3SiO2 } \rightarrow \ce{2Al2O3 + 3Si}
\end{equation}

We notice also a considerable presence of SiOx at the top surface of the metal. The source of this is likely atmospheric.

\begin{figure} [b]
    \centering
    \includegraphics[width=8.5cm]{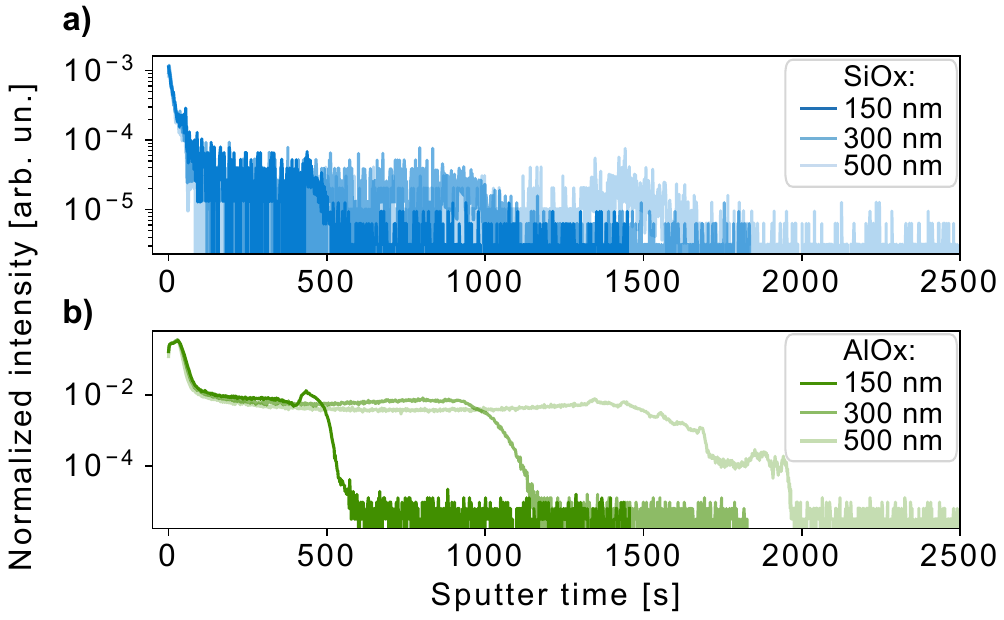}
    \caption{\label{fig:SIMS_supp} 
    Depth profiles of \textbf{a)} silicon oxide and \textbf{b)} aluminum oxide in the samples examined in figure~\ref{fig:SIMS}.   }
\end{figure}

\begin{figure}
    \centering
    \includegraphics[width=8.5cm]{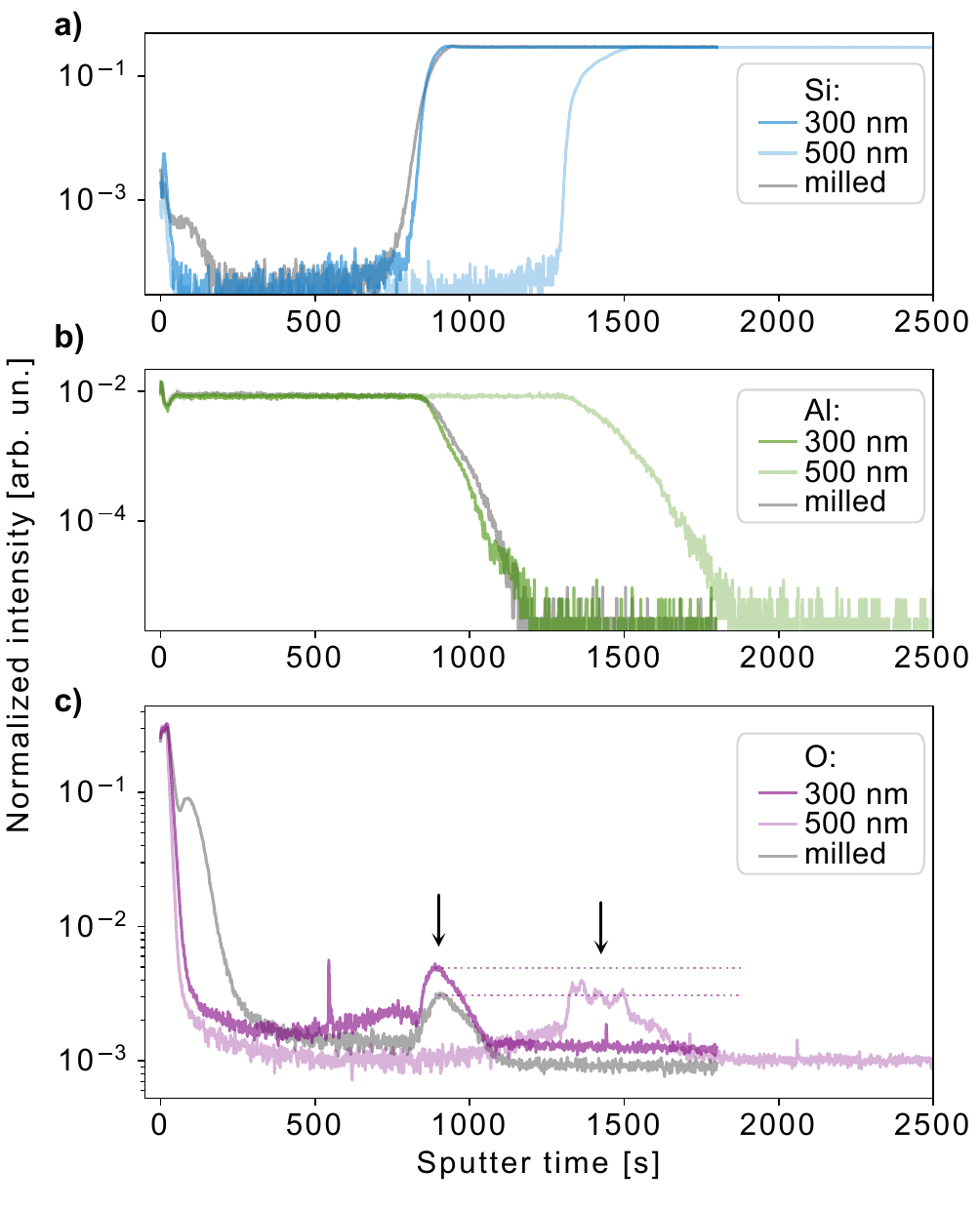}
    \caption{\label{fig:SIMS_milled} 
    Elemental analysis of the Al/Si material stack by ToF-SIMS. Depth profiles of \textbf{a)} Si, \textbf{b)} Al and \textbf{c)} O. Each plot contains data on three samples. The sample in grey shows an Al film milled from 500 nm to 300 nm, the other two samples are Al film of the two respective thicknesses. The sharp increase in the intensity of Si in \textbf{(a)}, and the arrows in \textbf{(c)}, mark the SM interface of each sample.}
\end{figure}

In addition, we analyze a sample where a 500 nm thick film was initially deposited, and later reduced down to 300 nm via argon ion milling. This is done to draw a comparison between this sample and the films of the two respective thicknesses, and trace the origin of the change in interfacial composition. These measurements were performed at 9.1 -- \SI{9.9e-9}{\milli\bar}. We present the obtained intensity traces for Si, Al and O for two samples of the representative thicknesses (\SI{300}{\nano\meter} and \SI{500}{\nano\meter}, same samples as figure~\ref{fig:SIMS}), as well as the sample reduced from \SI{500}{\nano\meter} to \SI{300}{\nano\meter}, in figure~\ref{fig:SIMS_milled}.

Figure~\ref{fig:SIMS_milled}(c) shows the O intensity. At the MA interface, the milled sample shows a higher O content in the top layers, likely due to damage during milling. At the SM interface, the milled sample does not suffer as strongly from decreased depth resolution as the 500 nm thick film, confirming that the SM peak broadening we see in thicker films is indeed a measurement artifact. On the other hand, the milled film retains a comparably low oxygen level at the SM interface as the \SI{500}{\nano\meter} film.

\begin{figure}[H]
    \centering
    \includegraphics[width=6.5cm]{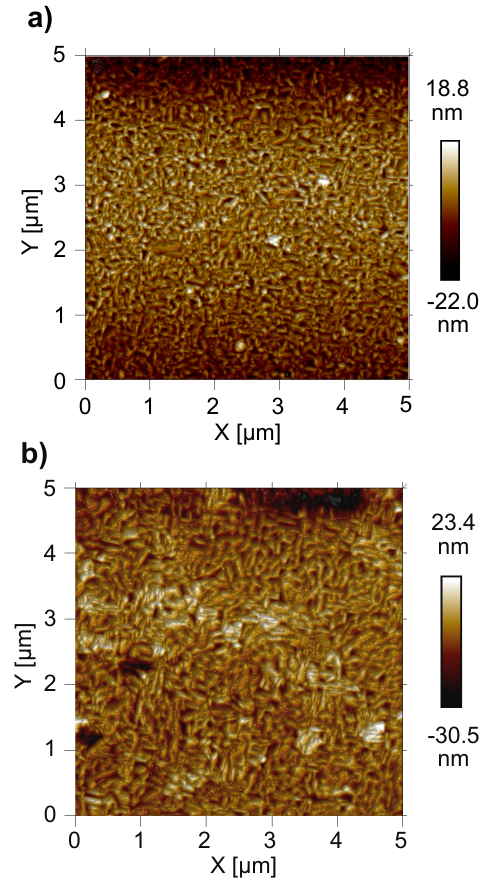}
    \caption{\label{fig:AFM} 
    AFM of the aluminum film surface. \textbf{a)} 150nm, and \textbf{b)} 500nm thick film.}
\end{figure}

\vfill

\section*{}

\bibliography{refs.bib}

\end{document}